\documentclass[final,5p,times,twocolumn]{elsarticle}
\usepackage{amssymb}
\usepackage{booktabs} % Horizontal rules in tables
\usepackage{multirow}
\usepackage{siunitx}
\usepackage{textcomp}
\usepackage{subfigure}
\usepackage{stfloats}
\usepackage{graphicx}
\journal{Nuclear Physics A}

\usepackage{float}
\usepackage{amsmath}

\begin{document}
\bibliographystyle{unsrt}
\begin{frontmatter}


\title{Title\tnoteref{label1}}
\tnotetext[label1]{Work supported by China Scholarship Council}

\author[LAL,IFP]{K.Wang\corref{mycorrespondingauthor}}
\cortext[mycorrespondingauthor]{Corresponding author}
\ead{wang@lal.in2p3.fr}
\author[CLUPS]{E.Baynard}
\author[LAL]{C.Bruni}
\author[LAL]{K.Cassou}
\author[LAL]{V.Chaumat}
\author[LAL]{N.Delerue}
\author[LPGP]{J.Demailly}
\author[LAL]{D.Douillet}
\author[LAL]{N.El.Kamchi}
\author[LIDYL]{D.Garzella}
\author[LPGP]{O.Guilbaud}
\author[LAL]{S.Jenzer}
\author[LPGP]{S.Kazamias}
\author[LAL]{V.Kubytskyi}
\author[LAL]{P.Lepercq}
\author[LPGP]{B.Lucas}
\author[LPGP]{G.Maynard}
\author[LPGP]{O.Neveu}
\author[CLUPS]{M.Pittman}
\author[CLIO]{R.Prazeres}
\author[LAL]{H.Purwar}
\author[LPGP]{D.Ros}

 \address[LAL]{LAL, Univ. Paris-Sud, CNRS/IN2P3, Universit\'e Paris-Saclay, Orsay, France.}
 \address[CLUPS]{CLUPS, Univ. Paris-Sud, Universit\'e Paris-Saclay, Orsay, France.}
 \address[LPGP]{Laboratoire de Physique des Gaz et des Plasmas, Univ. Paris-Sud, CNRS, Universit\'e Paris-Saclay, Orsay, France.}
 \address[CLIO]{CLIO/LCP, Univ. Paris-Sud, CNRS, Universit\'e Paris-Saclay, Orsay, France.}
 \address[LIDYL]{CEA/DRF/LIDYL, Universit\'e Paris-Saclay, Saclay, France.}
\address[IFP]{Institute of Fluid Physics, China Academy of Engineering Physics, P.O. Box 919-106, Mianyang 621900, China}

%%\fntext[label2]{}
%%\cortext[cor1]{}
%%\address{Address\fnref{label3}}

\title{Longitudinal compression and transverse matching of electron bunch for external injection LPWA at ESCULAP}

%%\author[label1,label2]{ }
%\address[label2]{LAL, Univ. Paris-Sud, CNRS/IN2P3, Universit\'e Paris-Saclay, Orsay, France}
%\address[label3]{LCP/CLIO, Univ. Paris-Sud, CNRS, Universit\'e Paris-Saclay, Orsay, France}
%\address[label4]{Laboratoire de Physique des Gaz et des Plasmas,
%	 Univ Paris-Sud, CNRS, Universit\'e Paris-Saclay, Orsay, France}
% \address[label6]{CLUPS, Universit\'e Paris-Sud, Universit\'e Paris-Saclay, Orsay, France.}
%\address[label5]{Institute of Fluid Physics, China Academy of Engineering Physics, P.O. Box 919-106, Mianyang 621900, China}
\begin{abstract}
We present theoretical and numerical studies of longitudinal compression and transverse matching of electron bunch before injecting into the Laser-plasma Wake Field Accelerator (LWFA) foreseen at the ESCULAP project in ORSAY. Longitudinal compression is performed with a dogleg chicane, the chicane is designed based on theory of beam optics, beam dynamics in dogleg is studied with ImpactT \cite{qiang2006three} and cross checked with CSRtrack \cite{dohlus2004csrtrack}, both 3D space charge (SC) and coherent synchrotron radiation (CSR) effects are included. Simulation results show that the energy chirp at the dogleg entrance should be smaller than the nominal optic design value, in order to compensate the negative energy chirp increase caused by longitudinal SC, while CSR can be ignored in our case. With an optimized configuration, the electron bunch ($\sim$10MeV, 10pC) is compressed from 0.9ps RMS to 70fs RMS (53\si{fs} FWHM), with a peak current of 152A. Transverse matching is realized with a doublet and a triplet, they are matched with Madx and the electron bunch is tracked with ImpactT, simulation results show little difference with the nominal design values, that is due to the SC effect. Finally, by simply adjusting the quadrupole strength, a preliminary optimized configuration has been achieved, that matches the Courant-Snyder (C-S) parameters to $\alpha_x=0.01,\alpha_y=-0.02, \beta_x=0.014\si{m},\beta_y=0.012\si{m}$ at the plasma entrance. 
\end{abstract}


\end{frontmatter}


\section{Introduction}

Over the past decades, significant progress has been made in the LWFA research\cite{faure2004laser,leemans2006gev}, but beam stability and quality are still needed to be improved to meet the requirements of most applications. Conventional RF injector can provide a stable and well understood external electron beam for LWFA, thus it provides the possibility to optimize the subsequent beam dynamics in plasma. There are some running and planned projects worldwide aiming to study the RF external injection scheme\cite{assmann2014sinbad-full,dorda2016sinbad,walker2017horizon,rossi2014external,zeitler2016phase}.

The objective of the ESCULAP project at LAL-Orsay\cite{delerue2016simulations} is to inject a relativistic electron beam generated by the RF photo injector PHIL \cite{alves2013phil} into a plasma wave excited by the 40 TW Laser LASERIX \cite{ple2007design} to
perform external injection LPWA experiment in quasi linear region. The ~5 MeV, \SI{10}{PC} electron bunch generated by PHIL has a duration of $\sim$0.9ps RMS. In order to slow the rate of phase slippage \cite{zeitler2016phase} and get an efficient acceleration in plasma wakefield, the electron bunch is accelerated off crest to $\sim$\SI{10}{MeV} with a s-band booster, then compressed with a dogleg chicane. After the dogleg, a doublet and a triplet are used to match the C-S parameters to the focusing force in the plasma. 
%%However in this scheme the energy gain in photocathode is comparable with that in off crest acceleration section (booster), uncorrelated energy spread is relatively large, final bunch length may be limited by uncorrelated energy spread.\\

Generally, the design of the bunch compression chicane is to match the energy chirp with $R_{56}$, suppress the effects introduced by chromatic abberation and RF curvature. The design is firstly based on the theory of beam optics, but actually, space charge force may affect the energy chirp (i.e. due to the charge repulsion force, the bunch head may get energy while the bunch tail lose energy, then induce a negative energy chirp, here, the convention is used that the leading ‘head’ of the bunch is in the direction $z<0$). In our case, dogleg has a positive $R_{56}$, a negative energy chirp $k_1<0$ is required, then the space charge may increase the energy chirp and lead to over compression, this is verified by our simulation. There are also some studies on producing a negative energy chirp with space charge instead of RF structure for bunch compression \cite{he2015design}.

 After compression, C-S parameters of the electron bunch should be matched to the plasma wakefield, a mismatched beam will experience significant emittance growth in the plasma. The emittance growth can be explained using a theory that particles of various energy experience different focus strength, thus correlated energy spread lead to varying betatron oscillation frequency between beam slices, while uncorrelated energy spread generates a betatron decoherence within the beam slices, as a result, they generate both slice and total emittance increase. A matched beam satisfies the condition that $\alpha_{x,y}=0, \beta_{x,y}=1/\sqrt{\overline{K}}=1/\overline{k_{\beta}}$, $\overline{K}$ is the mean focus strength exerted on electrons, and $\overline{k_{\beta}}$ is the mean betatron oscillation wave number of electrons in the plasma \cite{mehrling2014theoretical}.\\
In this paper, we present theoretical and numerical studies of longitudinal bunch compression and transverse bunch matching for the ESCULAP project. 
\section{Bunch compression}
\label{BC} 
\subsection{Dogleg design}
Magnetic bunch compression is based on the principle that when passing through a bend section, electrons of different energy travel different trajectories, then the bunch length can be controlled by  appropriately matching the dispersive parameters of the bend section and the correlated energy spread of the electron bunch, the correlated energy spread can be introduced by off crest acceleration. By looking at the ESCULAP beam line in figure 1, we simply review the theory of bunch compression,  assume that the reference particle has an energy of $E_{i0}$ after the RF gun, and then is accelerated to the nominal energy $E_{f0}$  at an RF phase $\phi_{0}$. For a general electron which is at longitudinal position $z_{i}$ with respect to reference particle, it has an energy of $E_{i}$ after RF gun and then be accelerated to $E_{f}$ 
\begin{equation}
E_{f}=E_{i}+{E_{f0}-E_{i0} \over cos(\phi_{0})}cos(\phi_{0}+2\pi z_{i}/\lambda)
\label{eq:1}
\end{equation}
$\lambda$ is the wavelength of the acceleration field in RF structure, generally, $z_{i} \ll \lambda/2 \pi$ and the relative energy deviation can be expanded as
\begin{equation}
\delta_{f} \approx \delta_{i}-(1-{E_{i0} \over E_{f0}}){2\pi \over\lambda}z_{i}tan(\phi_{0})-(1-{E_{i0} \over E_{f0}}){1 \over 2}({2\pi \over\lambda}z_{i})^2+(1-{E_{i0} \over E_{f0}}){1 \over 6}({2\pi \over\lambda}z_{i})^3tan(\phi_{0})
\label{eq:2}
\end{equation}
Here $\delta_{f}=(E_{f}-E_{f0})/E_{f0}, \delta_{i}=(E_{i}-E_{i0})/E_{f0}$, energy spread are both normalized with the final reference particle energy. After bend section, reference position of a general particle can be written as
\begin{equation}
z_{ex} \approx z_{i}+R_{56}\delta_{f}+T_{566}\delta_{f}^2+U_{5666}\delta_{f}^3+...
\label{eq:3}
\end{equation}
Submit equation \ref{eq:2} into equation \ref{eq:3}, recognise the coefficient of $z_i^m$ in equation \ref{eq:2} as $k_m$, and only persist items with $z_i^m,\delta_{i}^m, m<3$, then we have  
\begin{align}
z_{ex}^2 = & \quad R_{56}\delta_{i}^2 \nonumber \\ 
+ & \quad 2R_{56}\delta_{i}(1+k_{1}R_{56})z_i \nonumber\\
+ & \quad [(1+k_{1}R_{56})^2+2R_{56}\delta_{i}(k_{2}R_{56}+k_{1}^2T_{566})]z_{i}^2 \nonumber\\      
%%+ & \quad ..
\end{align}
The RMS bunch length is 
\begin{equation}
\sigma_{zex} \approx \sqrt{\quad R_{56}^2\sigma_{\delta i}^2+[(1+k_{1}R_{56})^2+2R_{56}\delta_{i}(k_{2}R_{56}+k_{1}^2T_{566})]\sigma_{zi}^2}
\label{eq:4}
\end{equation}
According to equation \ref{eq:4}, minimum bunch length can be acquired with full compression condition
\begin{equation}
R_{56}=-1/k_{1}, \quad T_{566}=k_{2}R_{56}/{k_{1}^2}
\label{eq:5}
\end{equation}
Under full compression, the bunch length achievable by a chicane is $\sigma_{zex} \approx  R_{56}\sigma_{\delta i}$, which indicates that shorter bunch length can be achieved with weaker $R_{56}$ and smaller uncorrelated energy spread. Furthermore, when particles travel along a bend section, CSR may increase the energy spread and disturb the energy chirp, finally increase the bunch length, weak $R_{56}$ and large initial uncorrelated energy spread can suppress CSR. Thus to optimize the uncorrelated energy spread (larger or smaller) we need to check the strength of CSR effect, but a smaller $R_{56}$ is strongly desired. However, smaller $R_{56}$ calls for larger energy chirp $K_{1}=-1/R_{56}$, increase of chirp may reduce the energy gain for a given RF power. In ESCULAP,  the electron bunch is accelerated by a booster, and the final energy should be $\sim$ 10MeV, then $R_{56}$ can't be smaller than 3.2cm. The uncorrelated energy spread is  $\sigma_{\delta i}=1.9 \times 10^{-4}$ (\SI{1.9}{keV} for \SI{10}{MeV}), then the bunch duration introduced by un correlated energy spread is \SI{26.2}{fs}. However because the higher order chromatic items, the couple between transverse and longitudinal direction, space charge(SC) and CSR effects, the achieveable duration is longer than this value.  \\
   
\begin{figure}[H]
	\centering
	\includegraphics[width=1\linewidth]{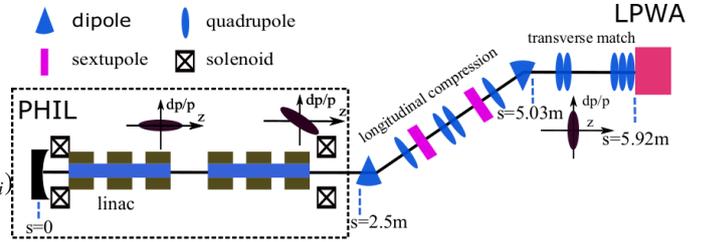}	
	\centering
	\caption{\textbf {\footnotesize \textit {Schematic of ESCULAP beam line}}}
	\label{fig:1}
	%	\smaller \textit{fig 1 the schematic of bunch compression line with dogleg}
\end{figure}

Emittance growth in the dogleg can be estimated with \cite{england2007longitudinal} $\Delta \epsilon=\sqrt{det(\epsilon_{0}+\eta \cdot \eta^T <\delta^2>+T \cdot T^T <\delta^4>)}$, $\eta=[R_{16}; R_{26}]$, $T=[T_{166}; T_{266}]$. 
The dogleg chicane in ESCULAP is designed with symmetry (figure 2), four quadrupoles are used to match $R_{16}=R_{26}=0$ and minimize the maximum beam size, two sextupoles are used to match $T_{566}$ to partially compensate the effects of RF curvature that my lead to fold over in phase space. And also a X-band cavity can be set before the chicane at deceleration phase to cancel the RF curvature \cite{emma2001x}, but it won't be used in ESCULAP. The matched result is $\rho =0.32\si{m}, \theta =\pi /4, k_1=39.33/\si{m},k2=18.35/\si{m},ks=614.53/\si{m^2} $, with the configuration that $L_{1}=0.45\si{m}, L_{2}=0.3\si{m}, L_{3}=0.12\si{m}$, $\rho$ and $\theta$ are respectively the radius and bend angle of dipoles. quadrupoles and sextupoles are with a thickness of $0.1\si{m}$. C-S parameters are shown in figure \ref{fig:3}, $R_{56}=4.49\si{cm}$.
\begin{figure}[H]
	\centering
	\includegraphics[width=6cm]{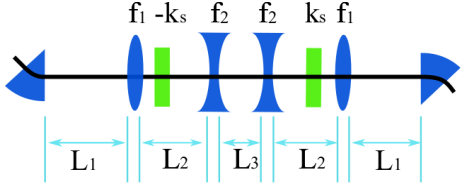}
	\caption{\textbf{ \footnotesize \textit {Schematic of the dogleg}}}
	\label{fig:2}
\end{figure}
\begin{figure}[H]
	\centering
	\includegraphics[width=6cm]{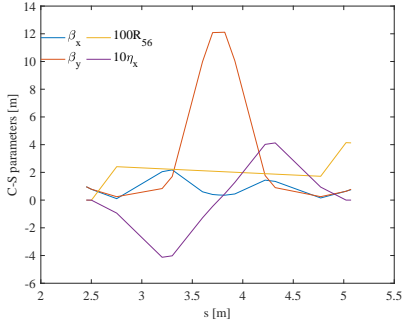}
	\caption{\textbf{ \footnotesize \textit {Nominal optic C-S parameters along the dogleg}}}
	\label{fig:3}
\end{figure}

\subsection{Simulation result}

\begin{figure*}[bp]
	\centering
	\subfigure[\textbf{ \footnotesize \textit{Electron bunch at the dogleg entrance}} ]{\includegraphics[width=6cm]{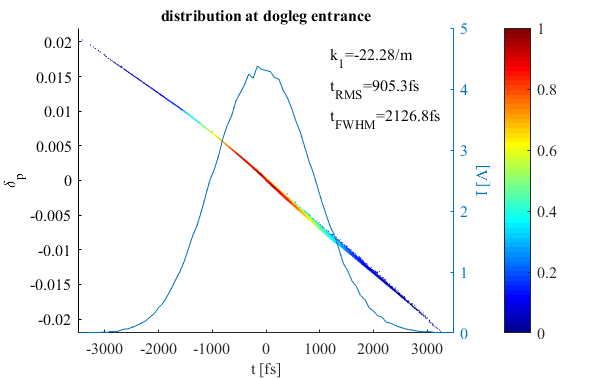}}
	\subfigure[\textbf{ \footnotesize \textit{Electron bunch at the dogleg exit}}] {\includegraphics[width=6cm]{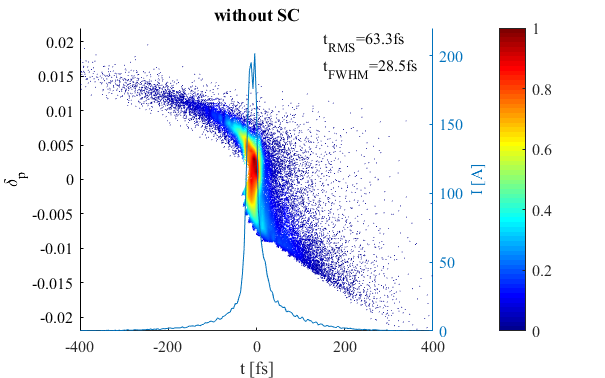}}
	\subfigure[\textbf{ \footnotesize \textit{Electron bunch at the dogleg exit}}] {\includegraphics[width=6cm]{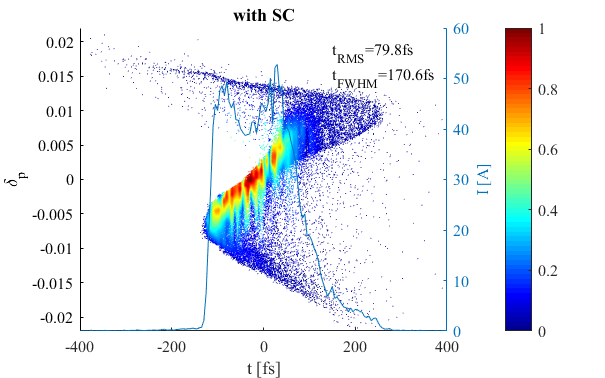}}
	\caption{\textbf{ \footnotesize \textit{Longitudinal distribution of electron bunch, a) at the dogleg entrance (s=2.5m), and at the dogleg exit (s=5.08m) tracked with ImpactT, b)without self-force, c)with space charge}}}
	\label{fig4}
\end{figure*}
\begin{figure*}[bp]
	\centering
	\subfigure[\textbf{ \footnotesize \textit{Tracked with ImpactT}}] {\includegraphics[width=6cm]{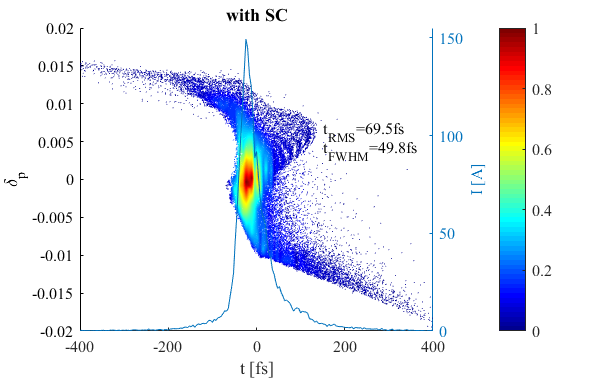}}
	\subfigure[\textbf{ \footnotesize \textit{Tracked with ImpactT}}] {\includegraphics[width=6cm]{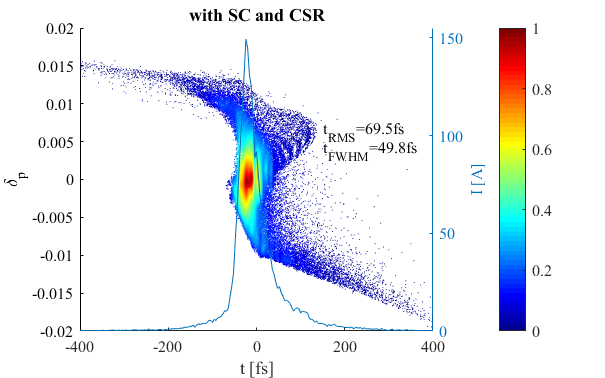}}
	\subfigure[\textbf{ \footnotesize \textit{Tracked with CSRtrack}}] {\includegraphics[width=6cm]{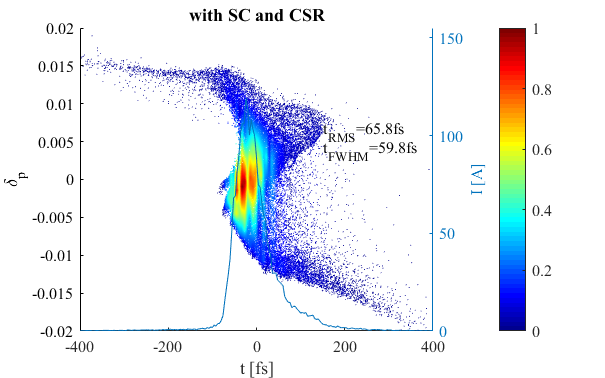}}
	\caption{ \textbf{ \footnotesize \textit{Longitudinal distribution at the dogleg exit (s=5.08m), a). Simulated with ImpactT without CSR; b) Simulated with ImpactT with both SC and CSR; c)Simulated with CSRtrack, csr$\_$g$\_$ to$\_$p force model is used, with $\sigma_{z.sub}=0.05\sigma_{z.local}$, $\sigma_{r.sub}=0.01$mm. Energy chirp at the dogleg entrance (s=2.5m) is $k_1=21.17/\si{m}$  }}}
	\label{fig5}
\end{figure*}
The electron bunch is tracked from cathode to the dogleg entrance with ASTRA \cite{flottmann2011astra} (50000 macro particles are used), the nominal energy chirp needed for full compression is $k_1=-1/R_{56}=-22.28/\si{m}$, this can be acquired by adjusting the phase and energy of the booster according to equation \ref{eq:2}. The longitudinal phase space at dogleg entrance is shown in figure 4a, the energy chirp is calculated with $k_1=<\delta \cdot z>/<z^2>$. In the dogleg, the electron bunch is tracked with ImpactT, either with SC and CSR turned off or with SC turned on, to see the effect of SC, the tracking result is shown in figure 4b and 4c respectively. Without space charge the electron bunch is compressed to 63.3fs RMS and 28.5fs FWHM, with a peak current of $\sim200$A.\footnote{After compression, The longitudinal distribution has a sharp peak, it's not Gaussian, FWHM value is more meaningful for the long tail.} However, with SC included, the bunch is over compressed and has a flat top, the peak current falls to $\sim$58A.

The difference between figure 4b and 4c is only caused by SC, SC increases the energy chirp and leads to over compression, this can be compensated by decreasing the initial energy chip at the dogleg entrance. An optimized result is shown in figure 5a, the energy chirp at the dogleg entrance is $K_1=22.17/\si{mm}$, after compression, electron bunch has a duration of 69.5fs  RMS and 53.2fs FWHM, with a peak current 150A. In order to check the effects of CSR, the electron bunch is tracked again with both SC and CSR turned on, the result is shown in figure 5b, CSR seems to has no effect on the longitudinal distribution. The simulation result is cross-checked with CSRtrack (csr$\_$g$\_$ to$\_$p method, with $\sigma_{z.sub}=0.05\sigma_{z.local}$, $\sigma_{r.sub}=0.01$mm) and the result is shown in figure 5c, the longitudinal phase spaces obtained by CSRtrack and ImpactT are in good agreement, while the little difference in bunch duration and transverse size may be caused by the differences in magnets models and simulation models between ImpactT and CSRtrack, a detailed discussion can be found in \cite{zhu2016sub}.

According to the simulation results, both FWHM and RMS bunch durations are smaller than 100fs with SC and CSR included, the final bunch length is dominated by uncorrelated energy spread, residual chromatic items, and SC effects. As analysed before, uncorrelated energy spread can induce a bunch duration of $\sim$26.2fs, this can be suppressed by designing a dogleg with smaller $R_{56}$, and decrease the uncorrelated energy spread. However, smaller $R_{56}$ require larger energy chirp to match, then longer RF structure is needed to get same acceleration. The uncorrelated energy spread can be decreased by optimizing the photo cathode \cite{huang2005uncorrelated} or increasing the energy gain in off crest acceleration.  \\
Transverse emittance growth and bunch size after the compression is shown table I, the initial emittance and bunch size are $\epsilon_{nx0}=\epsilon_{ny0}=0.62 \si{\mu m \cdot rad}, \sigma_x=\sigma_y=0.22\si{mm}$.  According to our simulation result with ImpactT, SC leads to significant emittance growth ($\sim76\%$) in x direction (bend plane), while only $4.7\%$ in y direction. The effects or CSR in X and Y direction are both smaller than $4\%$, that can be ignored.

\begin{center}
	Table I \textbf{\footnotesize \textit {Emittance growth and bunch size after compression (s=5.08m)}}\\
	\begin{tabular}{ccccccc}
		\toprule
		code & SC & CSR & $\frac{\Delta \epsilon_{nx}}{\epsilon_{nx0}} $  & $ \frac{\Delta \epsilon_{ny}}{\epsilon_{ny0}}$ &  $\sigma_{x} (\si{mm})$ & $\sigma_{y} (\si{mm})$  \\
%		\midrule
%		bf dogleg & 0.62 & $\frac{2.1}{1.2}$ & 0.31 & 0.31 \\
%		\midrule
		ImpactT & off & off & 3.65\% & 4.73\% & 0.23 & 0.22 \\
		ImpactT & on & off & 79.40\% & 7.31\% & 0.29 & 0.21\\
        ImpactT & on & on & 82.61\% & 4.42\% & 0.30 & 0.21\\
        CSRtrack & on & on & 85.13\% & 4.62\% & 0.30 & 0.21\\
		\bottomrule 
	\end{tabular}
\end{center} 


%%\bibliographystyle{elsarticle-num} 

\section{Transverse matching}
After the dogleg, the electron bunch has a relatively large transverse size ($\sim$0.29mm), before injecting into the plasma, it needs to be focused, and further more, the betatron function needs to be matched to the focusing field in the plasma. In our configuration, the electron bunch will be accelerated in quasi linear region in the plasma, and a preliminary simulation has been done, it shows that the average betatron oscillation has a duration of about 1cm in uniform density of $2 \times 10^{17} /\si{cm^3}$, which calls for a matched condition: $\alpha_{x,y}=0,  \beta_{x,y}=0.0016\si{m}$. However, a up ramp density profile and also a focusing laser\cite{dornmair2015emittance,xu2016physics} is to be used for transverse matching before acceleration. Here we match the electron bunch to $\alpha_{x,y}=0, \beta_{x,y}=0.01\si{m}$ by using a doublet and a triplet. \\
\begin{figure}[H]
	\centering
	\includegraphics[width=6cm]{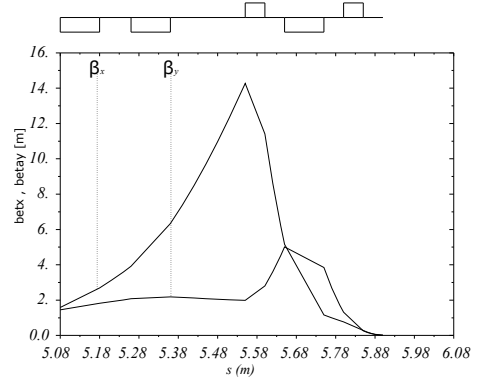}
	\caption{\textbf{ \footnotesize \textit {Nominal optic betatron function along match line}}}
	\label{fig:4}
\end{figure}

The matching is first realized with Madx, matching result is shown in figure 6. Emittance growth introduced by chromatic aberration can be estimated with 
$\epsilon_{x}^2=det[<Rxx^TR^T+Txx^TT\delta ^2] $, here $ T=[T_{116},T_{126};T_{216},T_{226}]$, and refer to the result in Madx, emittance has a growth of $\sim$20.1\% in x direction and $\sim$24.3\% in y direction.
However, the electron bunch is fully compressed, the actual betatron fuction may be affected by SC. The simulation result obtained by ImpactT (SC included) is $\beta_x=0.032\si{m}, \beta_y=0.028\si{m}$. Here we simply  scan the quadrupole strength around the nominal value(the optic design value in Madx) to minimize the mismatch factor\cite{minty1999beam} $B_{mag}=\frac{1}{2}(\beta \gamma_0 -2\alpha \alpha_0 +\gamma \beta_0)$, $\beta_0, \alpha_0, \gamma_0$ are the design parameters and $\beta, \alpha, \gamma$ are from the simulation results. Finally we get a preliminary optimized solution shown in table II  
\begin{center}
	Table II \textbf{\footnotesize \textit {Bunch parameters at the plasma entrance (s=5.92m)}}\\
	\begin{tabular}{cccccccccccc}
		\toprule
		para & nominal   & simulation &  optimized &  \\
		%		\midrule
		%		bf dogleg & 0.62 & $\frac{2.1}{1.2}$ & 0.31 & 0.31 \\
		\midrule
		$\alpha_x $ & 0.000 & 1.164 & 0.001  \\
		$\alpha_y$  & 0.000 & 1.063 & -0.002  \\
		$\beta_x$ [m]& 0.010 & 0.032 & 0.014  \\
		$\beta_y$ [m]& 0.010 & 0.028 & 0.012  \\
		$\sigma_x $[mm] & - & 0.051 & 0.047  \\
		$\sigma_y $[mm] & - & 0.037 &  0.035 \\
		$t_{RMS}$ [fs] & - & 110.2 & 113.3 \\
		$t_{FWHM}$ [fs] & - & 87.3 & 87.3 \\
		$B_{mag.x}$        & - & 1.981 & 1.057 \\
		$B_{mag.y}$        & - & 1.777 & 1.017 \\
		\bottomrule 
	\end{tabular}
\end{center}

To get the preliminary optimized result, the strength of five quadrupoles are increased respectively by 0.3\%, 0.9\%, 1.6\%, 1.2\%, 1.9\% , all smaller than 2\%, however the procedure is time consuming. The final RMS bunch duration is 113.3fs and FWHM bunch duration is 87.3fs, all the parameters satisfy the requirement of ESCULAP. Such a small FWHM value, compared to the RMS one, indicates the presence of a sharp peak at the maximum   of the longitudinal distribution, as also seen in Fig.5, and which is beneficial for an optimal coupling to the plasma wave. However, to suppress the bunch duration growth in the matching section, the electron bunch at the dogleg exit should be further optimized (i.e. an over compressed bunch may slightly compensate the bunch duration growth).   

\section{Summary} 
In this paper, we have presented theoretical and numerical studies of longitudinal compression and transverse matching of an external electron bunch before injecting into the LPWA at ESCULAP. By using a dogleg chicane, the electron bunch ($\sim$\SI{10}{MeV}) is compressed from \SI{0.9}{ps} RMS to \SI{70}{fs} RMS (\SI{53}{fs} FWHM), the final bunch duration is mainly limited by uncorrelated energy spread, chromatic abberation, and SC. It is noteworthy that SC can increase the energy chirp and lead to over compression, this can be compensated by decreasing the energy chirp at the dogleg entrance. Uncorrelated energy spread can be suppressed by optimizing the photocathode or accelerating the electron bunch off crest to higher energy, but this require longer RF structure. In our case, CSR has almost no effect on final bunch parameters. After the dogleg, a doublet and a triplet are used to match the electron bunch into the plasma. After the matching section, the C-S parameters are $\beta_x=0.014\si{m}, \beta_y=0.012\si{m}, \alpha_x=0.001 , \alpha_y= 0.002$. However, the bunch duration grows significantly in the matching section, to compensate it, an over compressed bunch is required after the dogleg, and further optimization is still necessary.


%%\bibitem{}
\bibliographystyle{elsarticle-num} 
%%\end{thebibliography}
\section{Reference}
\bibliography{bibfile}
\end{document}